# Experimental Investigation of Condensation Predictions for Dust-Enriched Systems

Gokce Ustunisik[a,*], Denton S. Ebel[a,b], David Walker[b,a], and Joseph S. Boesenberg[c,a]

[a]Department of Earth and Planetary Sciences, American Museum of Natural History, New York, NY, 10024-5192, U.S.A.
[b]Department of Earth and Environmental Sciences, Lamont Doherty Earth Observatory of Columbia University, Palisades, NY, 10964-8000, U.S.A.
[c]Department of Geological Sciences, Brown University, Providence, RI, 02912, U.S.A.
[*]**Corresponding author**: Gokce Ustunisik, gustunisik@amnh.org

**Abstract:** Condensation models describe the equilibrium distribution of elements between coexisting phases (mineral solid solutions, silicate liquid, and vapor) in a closed chemical system, where the vapor phase is always present, using equations of state of the phases involved at a fixed total pressure (< 1 bar) and temperature (T). The VAPORS code uses a CaO-MgO-$Al_2O_3$-$SiO_2$ (CMAS) liquid model at T above the stability field of olivine, and the MELTS thermodynamics algorithm at lower T. Quenched high-T crystal + liquid assemblages are preserved in meteorites as Type B Ca-, Al-rich inclusions (CAIs), and olivine-rich ferromagnesian chondrules. Experimental tests of compositional regions within 100K of the predicted T of olivine stability may clarify the nature of the phases present, the phase boundaries, and the partition of trace elements among these phases. Twenty-three Pt-loop equilibrium experiments in seven phase fields on twelve bulk compositions at specific T and dust enrichment factors tested the predicted stability fields of forsteritic olivine ($Mg_2SiO_4$), enstatite ($MgSiO_3$), Cr-bearing spinel ($MgAl_2O_4$), perovskite ($CaTiO_3$), melilite ($Ca_2Al_2SiO_7$ - $Ca_2Mg_2Si_2O_7$) and/or grossite ($CaAl_4O_7$) crystallizing from liquid. Experimental results for forsterite, enstatite, and grossite are in very good agreement with predictions, both in chemistry and phase abundances. On the other hand the stability of spinel with olivine, and stability of perovskite and gehlenite are quite different from predictions. Perovskite is absent in all experiments. Even at low oxygen fugacity (IW-3.4), the most $TiO_2$-rich experiments do not crystallize Al-, Ti-bearing calcic pyroxene. The stability of spinel and olivine together is limited to a smaller phase field than is predicted. The melilite stability field is much larger than predicted, indicating a deficiency of current liquid or melilite activity models. In that respect, these experiments contribute to improving the data for calibrating thermodynamic models including MELTS.

# 1. INTRODUCTION

Vapor-liquid interactions may be important in the formation and evolution of the earliest solids including chondrules and calcium-aluminum-rich inclusions (CAIs) in protoplanetary disk environments. The bulk major element and mineral chemistry of many CAIs and chondrules closely resembles that of the high temperature solid assemblages predicted by equilibrium calculations to condense out of a cooling high-temperature gas of solar composition (Lord, 1965; Larimer, 1967; Grossman, 1972; Yoneda and Grossman, 1995). Despite uncertainties and differences in thermodynamic data, results of such calculations agree in most respects (Ebel, 2006, his Fig 1).

It is thought that the midplane of the protoplanetary disk became enriched in previously condensed mineral dust. Ebel and Grossman (2000) assumed a carbonaceous chondrite composition dubbed "CI" as an appropriate dust analog. In this model, dust enrichment by a factor $d$ describes a vapor of solar composition ($d = 1$) to which $d$ atoms of elements condensable above ~500K, (e.g., Ti, Al, Fe, S) are added for each atom of Ti, Al, Fe, S, etc. in the solar composition. Each Ti brings two O, each Al 3/2 O, etc., but Fe, Ni, Co are added as metal or sulfide. The total *system* composition is defined by this factor, so in Fig. 1, vertical lines describe changing temperature (T) at fixed total pressure ($P^{tot}$) and total system composition (vapor + condensed phases). Even at $d = 100$ or $d = 1000$, these systems are still dominated by $H_2$, which does not condense. The condensation calculation is carried out from 2400K, with all elements in the vapor phase, to 1300K, where nearly all refractory elements have condensed, all at constant total system (vapor + condensate) composition (Fig. 1). As T is decreased, the bulk composition of the *condensate* assemblage changes continuously, as Al, Ca, Ti, Mg, Si, Fe, etc. condense from the vapor as oxides, silicates, or metal. At constant $d$, vapor plus condensates constitute a closed chemical system, where the vapor phase is always present, and thermodynamic equilibrium is maintained among all phases. Here we investigate only the predicted condensate assemblages for fixed $d$ and T.

The addition of liquid models to condensation calculations allowed prediction of the equilibrium distribution of elements in disk environments between coexisting phases (solids, liquid, and vapor), using internally consistent thermodynamic descriptions (equations of state) of the phases involved at a fixed $P^{tot}$ and temperature. Yoneda and Grossman (1995) used the CaO-MgO-$Al_2O_3$-$SiO_2$ (CMAS) liquid model of Berman (1983) to condense vapors of solar composition and enriched in a dust of ordinary chondrite composition at high T and $P^{tot} < 1$ bar. Ebel and Grossman (2000) developed the VAPORS code (Ebel et al., 2000; Ebel, 2006) to include both the MELTS thermodynamics algorithm of Ghiorso and Sack (1995) for multicomponent silicate liquids and solid solutions (Berman, 1988; e.g., Sack and Ghiorso, 1994), and the CMAS model of Berman (1983). Because the MELTS model (Ghiorso and Sack, 1995) cannot address $SiO_2$-depleted liquids, such as those that might produce Type B CAIs (Yoneda and Grossman, 1995), the VAPORS code uses the Berman (1983) model at temperatures above the stability field of olivine (Ebel and Grossman, 2000). However, the CMAS liquid model (Berman, 1983) does *not* include $TiO_2$, so perovskite (or hibonite) stability relative to liquid cannot be accurately predicted using that model. This is important in understanding CAIs, because many refractory lithophile trace elements (e.g., REE) are calculated to partition strongly into hibonite ($CaAl_{12}O_{19}$) or perovskite ($CaTiO_3$) (Boynton, 1975; Lodders, 2003), and are observed to occur in perovskite (Simon et al. 1999) and Al-, Ti-bearing calcic pyroxene (Simon et al. 1991). Here, we test the predictions of Ebel and Grossman (2000) for dust-enriched systems where silicate liquid is predicted to be stable (Fig. 1).

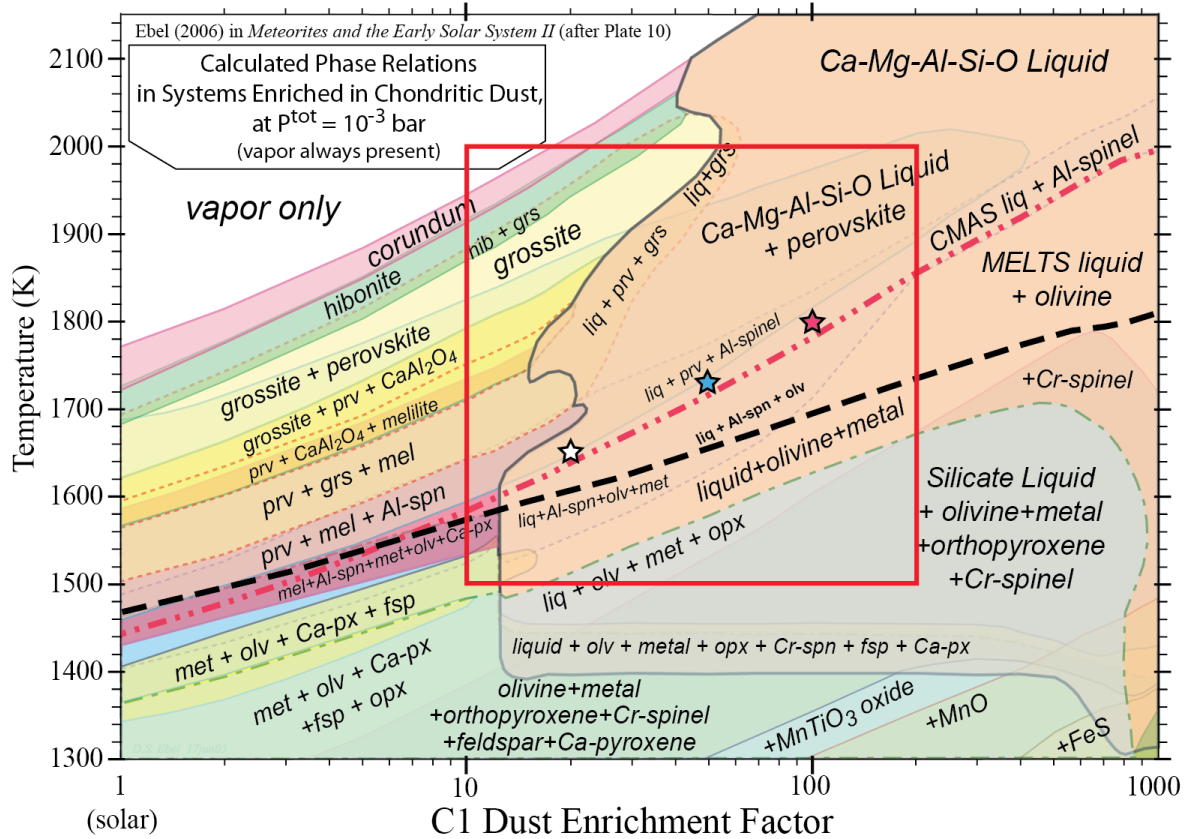

**Fig. 1.** Calculated phase relations during condensation in CI dust-enriched systems (modified from Ebel, 2006, plate 10). The red rectangle outlines the region targeted in experiments, with liquid predicted to be stable under most conditions (see Fig. 3). Stars mark bulk compositions and T referenced in Fig. 2.

Condensation from an $H_2$-rich vapor to an assemblage of solid(s) and/or liquid is controlled to first order by the difference in chemical free energy between vapor and condensates. In such a vapor, nearly all the Ca, Al, and Mg are present as the monatomic gas species, Ti is dominantly $TiO_{2(g)}$, and Si as $SiO_{(g)}$. For example, using data from Chase (1995), and Robie et al. (1978) for perovskite, the change in free energy for the reaction at 1800K and 0.1MPa (1 bar) for perovskite condensation from vapor

$$Ca_{(g)} + TiO_{2(g)} + O_{(g)} = CaTiO_{3(solid\ perovskite)} \qquad (1)$$

is -749.657 kJ./mole, and for

$$Ca_{(g)} + TiO_{2(g)} + O_{(g)} = CaO_{(liquid)} + TiO_{2(liquid)} \qquad (2)$$

the $\Delta G°_{reaction}$ is -569.524 kJ/mole. The free energy differences among the various condensed phases are small by comparison. The reaction

$$CaO_{(liquid)} + TiO_{2(liquid)} = CaTiO_{3(solid\ perovskite)} \qquad (3)$$

has $\Delta G°_{reaction}$ of -180.133 kJ/mole at 1800K. Predictions that some quantity of CaO, $TiO_2$, $Al_2O_3$, etc. will condense from a particular vapor at a particular T are, therefore, more likely to be reliable than predictions of exactly which phases will be stable, for example perovskite or $TiO_2$-bearing liquid. Stability fields among the condensates are highly dependent upon the thermochemical data and mixing properties used to describe minerals and liquids (Ebel, 2006).

Testing equilibrium between $H_2$-rich vapor and condensates in the laboratory is beyond the scope of this work, however, predictions of condensed liquid plus solid assemblages are amenable to testing by experiment. In this contribution, we perform crystallization experiments to systematically explore specific, narrow regions of temperature and total system composition space where either the CMAS model of Berman (1983) or the MELTS model of Ghiorso and Sack (1995) were applied by Ebel and Grossman (2000; cf., Ebel, 2006, his plate 10; Fig. 1). Each experiment tests whether the liquid + solid(s) assemblage that is calculated to be stable at a specific T and total system composition (dust enrichment), is the actual stable assemblage for the particular *condensate* bulk composition predicted to coexist with $H_2$-rich vapor.

These experiments also provide new calibration points to improve existing and future liquid activity models, data fundamental to the predictive infrastructure of geochemistry and cosmochemistry. They also may constrain the poorly known thermochemistry and mixing properties for minerals such as melilite, [$Ca_2Al_2SiO_7$ - $Ca_2(MgSi)SiO_7$] and Al-, Ti-rich calcic pyroxenes, perhaps [$Ca((MgSi),(Al,Ti^{3+}))(Ti^{4+}Si)O_6$], that are the most abundant high-temperature mineral solid solutions in many igneous CAIs.

## 2. METHODS

### 2.1. Experimental design

Each experimental bulk composition synthesized here represents the sum of the solid(s) and liquid condensates predicted to be in equilibrium with vapor at a specific T at $P^{tot} = 10^{-3}$ bar, for a system of solar composition enriched in CI-chondrite dust by a particular dust-enrichment factor ($d$; Ebel and Grossman, 2000). That is, experiments have the bulk condensate composition predicted to be in equilibrium with vapor at a specific T, for a specific initial total system composition. It is important to recognize that the predicted bulk condensate composition (solid(s) + liquid) at a specific $d$ factor will change with increase or decrease in temperature, since a portion of the total system will either evaporate or condense. Therefore, the bulk condensate composition at 1800 K for a vapor enriched at $d = 100$ differs from the bulk condensate composition at 1700 K and $d = 100$, because more Mg, Si, and O have condensed as olivine at the lower T (Fig. 1). In other words, in contrast to the usual phase diagram, not only the silicate liquid composition but also the bulk condensate composition, which is the bulk composition of each experiment, must change as the temperature decreases or increases. Therefore, each experiment represents a snapshot of the predicted condensed matter at a specific temperature and $d$, where $d$ defines the total *system* composition (vapor + condensed matter).

Boundaries of Al-spinel, olivine, and orthopyroxene phase stability fields in Fig. 1 have similar slopes, defining trend lines. This is illustrated in detail in Fig. 2 for $d = 20x$, $50x$, and $100x$, in the phase space at T just above olivine stability, where the predicted assemblage is CMAS liquid plus spinel (stars in Fig. 1). At all $d \leq 100$, more than 99% of the Ca, Ti, and Al are predicted to have condensed at the T of olivine condensation. Fig. 2 illustrates that bulk condensate compositions are nearly identical at different T and $d$ along this particular trend line (stars in Fig. 1), but also that bulk condensate compositions diverge more strongly at T far above and below olivine stability. Thus, it is valid to use one starting composition (yellow circles in Fig. 3) to probe phase stability at several T (white circles) that correspond to systems with different total composition (vapor plus condensate). This fortuitously facilitates the testing of multiple regions in the T-$d$ space by using one starting bulk composition within one phase assemblage boundary, at different T. In the case where a tested composition is in the middle of a large phase field where two end members are available as starting material, these compositions have been tested using both end members. For example, experiment #5 has been tested using both starting mix #1 (#5a) and #10 (#5b) at 1710K. Likewise, experiment #17 has been tested via both #4 (#17a) and #6 (#17b) at 1800K. In the case of #6 and #11 (half yellow circles), since the proxy bulk compositions #4 for #6 and #12 for #11 differ significantly in bulk composition due to the phase area being quite large, experiments have been tested by also making compositions #5 and #17 to directly test these specific T and $d$ conditions.

Systematic platinum loop experiments were designed using twelve bulk compositions (#1, #4, #6, #7, #10, #11, #12, #14, #15, #16, #18, and #23 of Table 1) to test equilibrium liquid plus solid assemblages that are calculated to be stable at a specific T, cooled from a vapor enriched by a specific $d$, for twenty three points in seven phase fields of T-$d$ space (Fig. 3a). Six predicted assemblages were tested only by proxy (Table 2; Fig. 3a, white circles). The range chosen was designed to test a critical region of condensation prediction space, as demonstrated by the very strong temperature dependence of bulk condensate compositions (Fig. 2). In this range, the fraction of total atoms predicted to condense from $H_2$-rich vapor increases, for example from $7.14 \times 10^{-4}$ to $6.05 \times 10^{-3}$ between 1800 and 1650 K at $d = 50x$. The range of predicted coexisting mineral phases changes from CAI assemblages to chondrule Mg-silicate assemblages in the range covered by these experiments.

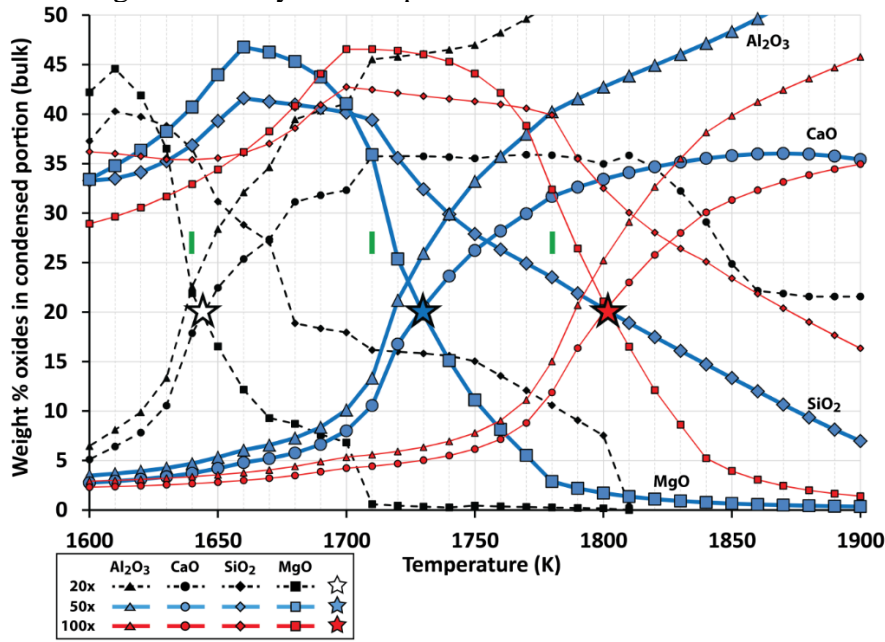

**Fig. 2.** Predicted bulk condensate compositions (liquid + solids) for vapors of solar composition enriched 20x, 50x, and 100x in CI-chondrite dust, neglecting Fe-Ni metal alloy. At each enrichment, a star indicates where MgO and CaO are equal (cross), an arbitrary criterion. The bulk condensate compositions at these crossings are strikingly similar, although their T differ greatly. In Fig. 1, these T and compositions are marked. These crossings (stars) move progressively above the T of olivine stability (marked with green line segments), with increasing dust enrichment.

## 2.2. Experimental details

Target bulk compositions of Table 1 were produced from mixtures of oxides, silicates, and carbonates by homogenizing the mixtures in ethanol in an automated mortar for >1 hour and drying at 175°C under vacuum to remove any adsorbed water. In the case of very high Ca/Si ratio bulk compositions (#1, #4, and #7), carbonate ($CaCO_3$) was used instead of silicate ($CaSiO_3$) as a source of CaO. For decarbonization of these compositions, the homogenized mixture was fired in a platinum crucible in air in a Deltech box furnace overnight by slowly heating the mixture to 1400°C via several steps (Step 1 to 810°C at 25°C/min, Step 2 to 920°C at 1°C/min, and Step 3 to 1400°C at 25°C/min). All starting compositions were prepared at the Department of Geosciences at Stony Brook University.

**Table 1.** Nominal starting compositions (wt%) used for Pt- loop equilibrium crystallization experiments.

| Oxide | #23 | #1 | #4 | #6 | #7 | #10 | #11 | #12 | #14 | #15 | #16 | #18 |
|---|---|---|---|---|---|---|---|---|---|---|---|---|
| $SiO_2$ | 34.01 | 26.33 | 23.41 | 19.13 | 13.57 | 28.81 | 39.73 | 40.61 | 49.84 | 44.35 | 52.09 | 5.09 |
| $TiO_2$ | 1.17 | 1.52 | 1.5 | 1.86 | 2.05 | 1.41 | 0.44 | 0.37 | 0.19 | 0.22 | 0.17 | 1.89 |
| $Al_2O_3$ | 26.18 | 35.75 | 39.82 | 43.81 | 48.22 | 32.09 | 9.89 | 8.36 | 4.22 | 4.87 | 3.91 | 59.03 |
| FeO | 0.02 | 0 | 0 | 0 | 0 | 0.02 | 0.12 | 0.2 | 0.7 | 0.66 | 1.25 | 0 |
| MgO | 17.58 | 8.14 | 3.95 | 1.08 | 0.37 | 12.17 | 41.9 | 43.75 | 41.66 | 46.05 | 38.94 | 0.18 |
| CaO | 20.83 | 28.2 | 31.32 | 34.12 | 35.79 | 25.36 | 7.83 | 6.62 | 3.35 | 3.86 | 3.09 | 33.81 |
| $Cr_2O_3$ | 0.21 | 0.06 | 0 | 0 | 0 | 0.14 | 0.1 | 0.09 | 0.04 | 0.01 | 0.54 | 0 |
| Total | 100 | 100 | 100 | 100 | 100 | 100 | 100 | 100 | 100 | 100 | 100 | 100 |

**Table 2.** Predicted bulk condensate compositions (wt%) tested using nominal compositions of Table 1 as proxies.

| Oxide | #5 | #8 | #9 | #17 | #20 | #22 |
|---|---|---|---|---|---|---|
| $SiO_2$ | 26.59 | 39.97 | 41.28 | 20.37 | 44.33 | 52.57 |
| $TiO_2$ | 1.55 | 0.4 | 0.34 | 1.75 | 0.22 | 0.17 |
| $Al_2O_3$ | 35.61 | 9.31 | 7.79 | 42.75 | 4.91 | 3.94 |
| FeO | 0 | 0.3 | 0.27 | 0 | 0.3 | 0.66 |
| MgO | 8.15 | 42.72 | 44.1 | 1.71 | 46.36 | 39.26 |
| CaO | 28.11 | 7.2 | 6.16 | 33.41 | 3.88 | 3.12 |
| $Cr_2O_3$ | 0 | 0.1 | 0.05 | 0 | 0 | 0.37 |
| Total | 100 | 100 | 100 | 100 | 100 | 100 |

In order to compare the compositions of prepared powders with the target compositions, an experimental glass was synthesized from #23 by melting this powder in a graphite capsule which was inserted, with dried pyrophyllite spacers, into a graphite furnace, which, in turn, was inserted into a talc sleeve. The sample was pressurized in a piston-cylinder apparatus at 1 GPa and heated to 1650°C for ~1 hour to ensure complete melting. This synthesis experiment produced a mostly homogeneous glass with tiny euhedral quench Mg-aluminate spinel crystals at the liquidus, concentrated near the walls of the capsule material. Since the quench spinels were too small to analyze but the rest of the glass was in quite good agreement with the target #23 composition, the composition of starting glass synthesized at 1GPa was not included in Table 1 and other starting compositions were not exposed to further high pressure synthesis before the one atmosphere gas mixing experiments.

Crystallization experiments were conducted at Lamont-Doherty Earth Observatory of Columbia University. Powders of twelve starting materials were pressed into 2-4 mm diameter and 1-2 mm thick pellets using polyvinyl alcohol (PVA) and hung on Pt-wire loops. The use of thin (0.1 mm) Pt-wire, the extremely low FeO contents of the mixtures and fayalite-magnetite-quartz (FMQ) buffer conditions assure almost no Fe alloying with the Pt-wire. Each pellet was then hung at the hot spot of a one atmosphere Deltech vertical quenching gas mixing furnace. In the case of identical target temperatures for different starting compositions (e.g., #1, #6, and #7), multiple charges were hung simultaneously at the corners of horizontal 10-15 mm ceramic bars at the hot spot. The oxygen fugacity ($fO_2$) was controlled by mixing CO and $CO_2$ gas to produce an $fO_2$ at or near the FMQ buffer, and by an $H_2$-$CO_2$ mixture for experiments at low $fO_2$. Due to the very low FeO content of starting compositions, $fO_2$ was only expected to potentially affect the stability of $Ti^{3+}$-bearing calcic pyroxene (Beckett, 1986). Temperature was controlled by PID feedback using a $Pt_{94}Rh_6$-$Pt_{70}Rh_{30}$ (type B) thermocouple in the furnace cavity outside the gas-bearing tube. The temperature of the experiment was measured with a $Pt_{10}Rh$-Pt (type S) thermocouple within the solid electrolyte $pO_2$ sensor within the gas-bearing tube within the furnace, calibrated against the melting points of gold, lithium metasilicate, and diopside. Horizontal temperature differences at the hot spot were within 1°C. Each charge was heated to the desired temperature and held there for between 20 minutes and several days (Table 3) before quenching of the run by dropping the sample into cold water. No seed crystals were added to these experiments due to the fact that oxides nucleate and grow readily in $SiO_2$-depleted liquids. After experiments, run products were lightly crushed, mounted in epoxy and polished for optical and electron microprobe analysis. All experiments, their run conditions, predicted and experimental phase abundances (wt%), and final phase assemblages are listed in Table 3. To assess the effects of $fO_2$ on phase assemblages and abundances, compositions #1 and #6 were mounted and run at $fO_2$ =IW-3.4 (IW for iron-wüstite) using the same setup as above but with an $H_2$-$CO_2$ gas mixture.

## 2.3. Analytical techniques

Backscattered electron (BSE) images, X-ray intensity maps (Fig. 4), and mineral and glass compositions of experimental runs were obtained using the Cameca SX100 electron microprobe (EMP) at the American Museum of Natural History (AMNH). BSE images were used to determine textural characteristics in order to differentiate quench from equilibrium spinels. Mass balance calculations using the starting compositions of Table 1, the crystallizing phase compositions, and the residual glass compositions were done using the least squares algorithm of the IgPet software (Carr, 2000) in order to determine mineral abundances (wt%) and test that no phase was missed during microprobe analysis.

X-ray intensity maps were obtained to determine the presence or absence of phases, especially perovskite, which might be very small in size and not very abundant and therefore hard to spot in BSE images. Various portions of each run were mapped ($128^2$ or $256^2$ pixels) at 1 micron/pixel resolution, 15 kV accelerating voltage, and 20nA beam current using five wavelength dispersive spectrometers (WDS: Al, Fe, Ti, Mg, and Ca), an energy dispersive spectrometer (EDS: Al, Si, Ti, Cr, Fe), and a BSE detector. Instrument parameters were chosen to minimize acquisition time and optimize intensities. X-ray intensity maps were combined into composite red-green-blue (RGB) three element images to facilitate image analysis for particular phases (cf., Ebel et al., 2008).

Quantitative microprobe analysis conditions for minerals were: 15 kV accelerating voltage, 20 nA beam current, focused electron beam (nominally 1 µm), and peak and background counting times of 30-60 s. The same setup was used for the glass analyses, but with the electron beam at 10 µm except for the highly crystalline low temperature experiments where the electron beam for both glass and mineral analyses was kept at 1 µm. We analyzed for all elements in all experiments: Si, Ti, Al, Fe, Mg, Ca, and Cr. Analytical standards were well-characterized synthetic oxides and natural minerals (enstatite for Si, synthetic perovskite for Ca and Ti, synthetic corundum for Al, fayalite for Fe, synthetic $MgAl_2O_4$ spinel for Mg, and synthetic magnesiochromite for Cr). Tests on several secondary standards were performed throughout the analytical session to verify that calibrations did not drift. The average representative glass and mineral compositions for all experiments are shown in Table 4. Results for reduced experiments (IW-3.4) are given in Table 5.

## 3. RESULTS

Predicted liquid and solid assemblages in seven phase fields have been tested by a total of twenty-three experiments using twelve bulk compositions at specific temperatures. Phase presence or absence compared to the predicted assemblages is reported in Table 3. Mineral and glass compositions are presented in Table 4, with modal abundances (wt%) calculated using the IgPet software (Carr, 2000). Predicted and experimentally determined phase fields are shown in Fig. 3, with phase boundaries estimated.

Grossite ($CaAl_4O_7$) stability has been tested by #7 (1860K and 30x) and #18 (1760K and 20x). Both charges produced a homogeneous glass and euhedral to subhedral blocky and acicular grossite crystals. Modal abundances were calculated as 76% glass and 24% grossite in #7 and 84% glass and 16% grossite in #18. CaO and $Al_2O_3$ abundances of grossite are very similar in blocky and acicular crystals and also between the two experiments. Perovskite was not recovered in either of these experiments. Abundances of grossite (Table 3) were found to be three and eight times higher in experiments #18 and #7 respectively when compared to the abundances predicted by Ebel and Grossman (2000).

Predictions of perovskite ($CaTiO_3$) stability have also been tested by 5 experiments (#7, #17a, b, #4, #6) in a range of temperature and dust enrichments. None of these experiments grew perovskite. Bulk compositions #4 and #6 were sufficiently different that #17 was tested by using both compositions. Experiments #17a and #17b at 1800K and 50x stabilized quite different abundances of liquid (93% in #17a and 48% in #17b) and euhedral gehlenite crystals (7% in #17a and 52% in #17b). The chemistry of gehlenite as noted in Table 4 was quite similar with no characteristics of quench crystallization. As shown by the RGB images (Ti-Ca-Al map in Fig. 4a; see also the Ti x-ray intensity map Fig. 4b), Ti is slightly concentrated in glass at the rims of the crystals, which is an indication of the growth of gehlenite with Ti concentrated in adjacent liquid due to slow diffusion of Ti. Similar behavior can be observed in the Ti-Ca-Al map of #6-1 (Fig. 4f). Perovskite stability was also tested at 1850K and 100x (by #4) and 1760K and 30x (by #6-1 and #6-2). Experiment #4 produced homogenous glass with no crystals, while #6-1 produced 43% glass, 49% prismatic and blocky gehlenite with 8% euhedral to subhedral grossite. Experiment #6-2 (using mix #4 at 1760K) stabilized 92% glass with 5% blocky gehlenite crystals and 3% Al-spinel. Again, the bulk compositions of #4 and #6 are sufficiently different that the result of #6-2 was not considered in Fig 3b.

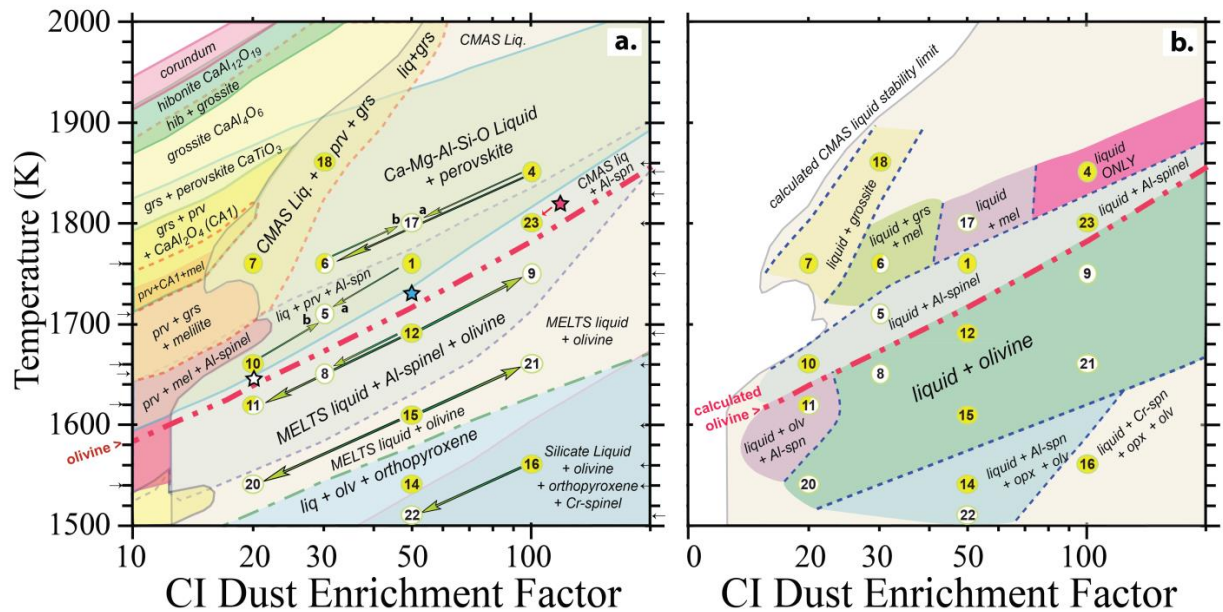

**Fig. 3.** Predicted (**a**) and experimentally determined (**b**) phase fields. Area is that outlined in Fig. 1. In (**b**), phase boundaries are estimated from experimental results. Twelve bulk compositions in yellow (#1, #4, #6, #7, #10, #11, #12, #14, #15, #16, #18, and #23) are used to test predicted liquid + solid assemblages at specific temperatures (T), cooled from a vapor enriched by a specific dust enrichment factor ($d$, e.g. 20x), for 23 points in T-$d$ space (**a**). Along trends indicated by arrows in (**a**), one bulk composition can be used as a proxy to test other compositions (white circle) in T-$d$ space (cf, Fig. 1). Experiment #5 has been tested using both #1 (#5a) and #10 (#5b). Likewise, experiment #17 has been tested via both #4 (#17a) and #6 (#17b). Experiment #6 (half yellow circle) was tested both using #6 as a starting composition and #4 as a proxy; likewise #11 used the exact predicted starting composition and also #12 as a proxy. Arrows on T-axes in (**a**) indicate T of experiments. Stars in (**a**) mark bulk compositions and T referenced in Figs 1 and 2.

Spinel stability has been confirmed by experiments #1, #5a, #5b, #10, and #23, while the predicted boundary between the CMAS liquid + Al-spinel + perovskite and CMAS liquid + Al-spinel fields was not found to exist. Although the bulk compositions of #1 (1760K at 50x) and #10 (1660K at 20x) are similar, the stability of phases at point #5 was tested using both bulk compositions. Experiments #1, #10, #5a and #5b (1710K at 30x) all produced homogeneous glass and 9 to 14% Al-spinel. Experiment #23 produced 93% glass and 7% Al-spinel which has ~3 wt% $Cr_2O_3$. Although experiments #10 and #5b have stabilized Al-spinel, the analyses are not included in Table 4 because the finer sizes of crystals made them impossible to analyze by EPMA without overlap with surrounding glass. The Mg-Ca-Al map of #5b (Fig. 4d) shows the concentration of spinel in various parts of the charge. Experimental abundances of spinel were always found to be lower than or equal to predicted abundances.

**Table 3.** Experimental conditions and results. Bold compositions are synthesized starting compositions. Predicted and found phases are liq: liquid, Al-spn: aluminum spinel, Cr-spn: chromium spinel, geh: gehlenite, grs: grossite, olv: olivine, prv: perovskite, opx: orthopyroxene. Spinels with * are too fine-grained to analyze. Gehlenite is not predicted and perovskite is not found under any conditions.

| Composition | Temperature (T, K) | Duration (t, min) | $d$ (xC1) | Predicted Abundances (wt %) | | | | | | | Experimental Abundances (wt %) | | | | | | Phases Found |
|---|---|---|---|---|---|---|---|---|---|---|---|---|---|---|---|---|---|
| | | | | prv | grs | spn | geh | olv | opx | liq | grs | spn | geh | olv | opx | gl | |
| **23** | 1800 | 60 | 100x | | | 11 | | | | 89 | | 7 | | | | 93 | gl +Al-spn |
| **4** | 1850 | 70 | 100x | 5 | | | | | | 95 | | | | | | 100 | gl |
| 6-2 (using 4) | 1760 | 73 | 30x | 6 | | | | | | 94 | | 3 | 5 | | | 92 | gl +Al-spn+geh |
| **6-1** | 1760 | 85 | 30x | 6 | | | | | | 94 | 8 | | 49 | | | 43 | gl +grs+ geh |
| 17a (using 4) | 1800 | 100 | 50x | 6 | | | | | | 94 | | | 7 | | | 93 | gl + geh |
| 17b (using 6) | 1800 | 55 | 50x | 6 | | | | | | 94 | | | 52 | | | 48 | gl + geh |
| **7** | 1760 | 54 | 20x | 7 | 3 | | | | | 90 | 24 | | | | | 76 | gl +grs |
| **1** | 1760 | 54 | 50x | 4 | | 12 | | | | 84 | | 9 | | | | 91 | gl +Al-spn |
| 5a (using 1) | 1710 | 90 | 30x | 5 | | 14 | | | | 81 | | 14 | | | | 86 | gl +Al-spn |
| 5b (using 10) | 1710 | 100 | 30x | 5 | | 14 | | | | 81 | | 1 | | | | 99 | gl +**Al-spn*** |
| **10** | 1660 | 60 | 20x | 3 | | 20 | | | | 77 | | 1 | | | | 99 | gl +**Al-spn*** |
| 9 (using 12) | 1750 | 90 | 100x | | 1 | | | 61 | | 38 | | | | 59 | | 41 | gl +olv |
| **12** | 1690 | 780 | 50x | | | 4 | | 63 | | 33 | | | | 61 | | 39 | gl +olv |
| 11-2 (using 12) | 1620 | 560 | 20x | | | 6 | | 60 | | 34 | | | | 67 | | 33 | gl +olv |
| 8 (using 12) | 1650 | 80 | 30x | | | 5 | | 62 | | 33 | | | | 68 | | 32 | gl +olv |
| **11-1** | 1620 | 1440 | 20x | | | 6 | | 60 | | 34 | | 5 | | 62 | | 33 | gl +Al-spn+olv |
| 20 (using 15) | 1540 | 2820 | 20x | | | | | 76 | | 24 | | | | 74 | | 26 | gl +olv |
| **15** | 1610 | 560 | 50x | | | | | 74 | | 26 | | | | 73 | | 27 | gl +olv |
| 21 (using 15) | 1660 | 2800 | 100x | | | | | 71 | | 29 | | | | 31 | | 69 | gl +olv |
| **14** | 1540 | 560 | 50x | | | | | 47 | 26 | 27 | | 1 | | 48 | 25 | 26 | gl +Al-spn+olv+opx |
| 22 (using 16) | 1510 | 3860 | 50x | | | 1 | | 39 | 36 | 24 | | 1 | | 50 | 23 | 26 | gl +Al-spn+olv+opx |
| **16** | 1540 | 560 | 100x | | | 1 | | 37 | 37 | 25 | | 1 | | 32 | 45 | 22 | gl +**Cr-spn***+ olv+opx |
| **18** | 1860 | 20 | 30x | 6 | 9 | | | | | 85 | 16 | | | | | 84 | gl +grs |

Al-spinel coexistence with olivine was tested using 5 experiments (#11-1, #11-2, #8, #12, and #9). Composition #12 has been used as a proxy for #8 (1650 K at 30x) and #9 (1750K at 100x). Both stabilized homogeneous glass (32-41 wt%) and euhedral forsteritic olivine (59-68 wt%), but no spinel crystallized in these experiments. Since the bulk composition of #11 differs significantly from #12, the phase assemblage of #11 was tested both using #12 as a proxy (#11-2), and as its own bulk composition (#11-1). Experiment #11-1 (at 1620K at 20x) is the only one found to crystallize both olivine (62%) and Al-spinel (5%) together. Experiment (#11-2) was not considered in Fig. 3b. Although the spinel composition reported in Table 3 contains high $SiO_2$ due to the electron beam overlapping with surrounding glass, Mg-Ca-Al maps clearly indicate that spinel grew in the experimental charge (Fig. 4e). Lack of spinel did not affect the good agreement between experimental and predicted abundances of olivine (Table 3; Fig 3).

Prediction of olivine stability without spinel was confirmed by experiments #20 (1540K at 20x), #15 (1610K at 50x), and #21 (1660K at 30x). All three experiments in this region produced 71-74 wt% abundances of forsteritic olivine similar in chemical composition, and similar to predicted olivine abundances (Ebel and Grossman 2000). Because #20 used composition #15 as a proxy (Fig. 3a), and considering that #11-1 crystallized Al-spinel but #11-2 (using composition #12 as proxy) did not, it is possible that Al-spinel could be stable at point #20 (Fig. 3b).

**Table 4.** Average residual glass and phase compositions (wt%) from Pt-loop equilibrium crystallization experiments. Standard deviations (1σ) for all values are given in italics in parentheses. Mg-, Al-spinel in #5b and #10, and Mg-, Cr-spinel in #16 were too small to analyze without overlap with adjacent $SiO_2$-bearing glass.

| Exp. | Phase/Abundance | $SiO_2$ | $TiO_2$ | $Al_2O_3$ | FeO | MgO | CaO | $Cr_2O_3$ | Total |
|---|---|---|---|---|---|---|---|---|---|
| #23 | Glass (93 wt%) | 38.28 *(0.61)* | 1.27 *(0.03)* | 22.59 *(0.82)* | 0.12 *(0.03)* | 16.04 *(0.41)* | 21.47 *(0.35)* | 0.04 *(0.01)* | 99.81 |
| | Spinel (7 %) | 0.58 *(0.44)* | 0.16 *(0.04)* | 67.70 *(1.79)* | 0.08 *(0.05)* | 27.36 *(0.68)* | 0.28 *(0.26)* | 2.88 *(0.74)* | 99.04 |
| #1 | Glass (91%) | 16.61 *(1.31)* | 0.90 *(0.08)* | 51.37 *(3.11)* | 0.00 *(0.00)* | 15.23 *(0.92)* | 15.74 *(1.66)* | 0.35 *(0.03)* | 100.2 |
| | Spinel (9%) | 0.12 *(0.26)* | 0.06 *(0.03)* | 70.81 *(0.18)* | 0.01 *(0.03)* | 28.26 *(0.41)* | 0.22 *(0.22)* | 0.70 *(0.50)* | 100.19 |
| #4 | Glass (100%) | 23.91 *(0.08)* | 1.46 *(0.02)* | 39.14 *(0.08)* | 0.01 *(0.01)* | 3.27 *(0.02)* | 31.47 *(0.09)* | 0.01 *(0.00)* | 99.6 |
| #5a (using 1) | Glass (86%) | 29.54 *(0.98)* | 1.64 *(0.24)* | 31.53 *(0.62)* | 0.01 *(0.01)* | 4.20 *(0.50)* | 32.31 *(1.39)* | 0.01 *(0.00)* | 99.23 |
| | Spinel (14%) | 0.71 *(0.70)* | 0.08 *(0.04)* | 70.34 *(0.99)* | 0.01 *(0.00)* | 27.00 *(0.99)* | 0.92 *(0.68)* | 0.53 *(0.02)* | 99.59 |
| #5b (using 10) | Glass (100%) | 34.11 *(0.09)* | 1.62 *(0.04)* | 28.15 *(0.06)* | 0.14 *(0.01)* | 8.46 *(0.05)* | 26.96 *(0.09)* | 0.01 *(0.00)* | 99.66 |
| #6-2 (using 4) | Glass (92%) | 23.72 *(0.17)* | 1.49 *(0.08)* | 39.61 *(0.42)* | 0.01 *(0.01)* | 3.04 *(0.23)* | 31.41 *(0.90)* | 0.00 *(0.00)* | 99.28 |
| | Spinel (3%) | 0.23 *(0.15)* | 0.04 *(0.00)* | 71.54 *(0.18)* | 0.01 *(0.00)* | 27.40 *(0.42)* | 0.51 *(0.19)* | 0.07 *(0.01)* | 99.79 |
| | Gehlenite (5%) | 21.63 *(0.03)* | 0.09 *(0.01)* | 36.35 *(0.05)* | 0.01 *(0.01)* | 0.48 *(0.01)* | 40.91 *(0.07)* | 0.00 *(0.00)* | 99.47 |
| #6-1 | Glass (43%) | 19.83 *(0.14)* | 4.02 *(0.01)* | 44.64 *(0.04)* | 0.01 *(0.01)* | 2.24 *(0.12)* | 27.68 *(0.39)* | 0.00 *(0.00)* | 98.41 |
| | Grossite (8%) | 0.71 *(0.14)* | 0.08 *(0.01)* | 75.92 *(0.13)* | 0.00 *(0.00)* | 0.31 *(0.04)* | 22.14 *(0.07)* | 0.01 *(0.01)* | 99.17 |
| | Gehlenite (49%) | 21.11 *(0.05)* | 0.21 *(0.04)* | 36.41 *(0.13)* | 0.00 *(0.00)* | 0.15 *(0.01)* | 41.00 *(0.12)* | 0.00 *(0.00)* | 98.89 |
| #7 | Glass (76%) | 17.78 *(1.37)* | 2.19 *(0.71)* | 39.70 *(1.19)* | 0.01 *(0.01)* | 0.30 *(0.08)* | 39.41 *(0.50)* | 0.00 *(0.00)* | 99.4 |
| | Grossite (24%) | 0.73 *(0.26)* | 0.24 *(0.06)* | 76.47 *(0.55)* | 0.01 *(0.01)* | 0.19 *(0.07)* | 22.27 *(0.41)* | 0.00 *(0.00)* | 99.92 |
| #8 | Glass (32%) | 46.90 *(0.17)* | 0.81 *(0.03)* | 19.47 *(0.26)* | 0.20 *(0.01)* | 18.61 *(0.67)* | 14.02 *(0.25)* | 0.08 *(0.00)* | 100.1 |
| | Olivine (68%) | 43.43 *(0.67)* | 0.03 *(0.03)* | 0.89 *(0.68)* | 0.15 *(0.01)* | 54.61 *(2.36)* | 0.86 *(0.50)* | 0.06 *(0.01)* | 100.04 |
| #9 | Glass (41%) | 45.55 *(0.63)* | 0.64 *(0.00)* | 16.25 *(0.42)* | 0.27 *(0.01)* | 24.76 *(0.83)* | 11.89 *(0.34)* | 0.08 *(0.00)* | 99.44 |
| | Olivine (59%) | 42.07 *(0.51)* | 0.04 *(0.02)* | 0.90 *(0.43)* | 0.18 *(0.01)* | 56.12 *(1.00)* | 0.84 *(0.30)* | 0.06 *(0.00)* | 100.21 |
| #10 | Glass (100%) | 36.48 *(0.21)* | 1.69 *(0.04)* | 25.14 *(0.18)* | 0.19 *(0.01)* | 6.96 *(0.04)* | 29.09 *(0.06)* | 0.01 *(0.01)* | 99.55 |
| #11-1 | Glass (33%) | 41.72 *(0.30)* | 1.55 *(0.05)* | 18.02 *(0.11)* | 0.22 *(0.01)* | 16.12 *(0.13)* | 21.07 *(0.09)* | 0.01 *(0.00)* | 98.7 |
| | Spinel (5%) | 4.30 *(1.31)* | 0.27 *(0.04)* | 65.58 *(3.44)* | 0.19 *(0.02)* | 26.80 *(0.49)* | 1.81 *(0.71)* | 1.36 *(0.29)* | 100.31 |
| | Olivine (62%) | 41.40 *(0.37)* | 0.03 *(0.01)* | 0.62 *(0.02)* | 0.17 *(0.01)* | 56.40 *(0.40)* | 1.01 *(0.21)* | 0.00 *(0.00)* | 99.64 |
| #11-2 (using 12) | Glass (33%) | 47.03 *(0.35)* | 0.85 *(0.03)* | 20.32 *(0.52)* | 0.14 *(0.01)* | 16.32 *(0.67)* | 14.89 *(0.21)* | 0.07 *(0.00)* | 99.63 |
| | Olivine (67%) | 43.10 *(0.55)* | 0.02 *(0.02)* | 0.26 *(0.02)* | 0.13 *(0.01)* | 56.94 *(0.97)* | 0.42 *(0.02)* | 0.06 *(0.00)* | 100.93 |
| #12 | Glass (39%) | 45.93 *(0.11)* | 0.75 *(0.00)* | 17.56 *(0.15)* | 0.14 *(0.00)* | 21.11 *(0.34)* | 12.96 *(0.02)* | 0.08 *(0.01)* | 98.53 |
| | Olivine (61%) | 41.61 *(0.47)* | 0.01 *(0.01)* | 0.17 *(0.07)* | 0.11 *(0.01)* | 56.52 *(0.48)* | 0.36 *(0.06)* | 0.06 *(0.00)* | 98.84 |

**Table 4** (*continued*)

| Exp. | Phase/Abundance | SiO$_2$ | TiO$_2$ | Al$_2$O$_3$ | FeO | MgO | CaO | Cr$_2$O$_3$ | Total |
|---|---|---|---|---|---|---|---|---|---|
| #14 | Glass (26%) | 54.45 *(0.03)* | 0.59 *(0.04)* | 14.72 *(0.30)* | 0.91 *(0.02)* | 17.61 *(3.09)* | 11.30 *(1.27)* | 0.02 *(0.00)* | 99.6 |
| | Olivine (48%) | 42.44 *(0.12)* | 0.02 *(0.00)* | 0.14 *(0.05)* | 0.09 *(0.02)* | 56.08 *(0.36)* | 0.36 *(0.06)* | 0.06 *(0.01)* | 99.2 |
| | Orthopyroxene (25%) | 58.55 *(0.28)* | 0.06 *(0.02)* | 0.87 *(0.17)* | 0.67 *(0.02)* | 38.60 *(0.26)* | 0.59 *(0.06)* | 0.03 *(0.01)* | 99.36 |
| | Al-Spinel (1%) | 0.31 *(0.06)* | 0.05 *(0.01)* | 74.34 *(0.59)* | 0.02 *(0.00)* | 24.00 *(0.39)* | 0.82 *(0.28)* | 0.73 *(0.02)* | 100.27 |
| #15 | Glass (27%) | 51.95 *(0.05)* | 0.75 *(0.02)* | 16.46 *(0.00)* | 0.58 *(0.01)* | 16.97 *(0.05)* | 12.09 *(0.02)* | 0.03 *(0.00)* | 98.34 |
| | Olivine (73%) | 42.02 *(0.09)* | 0.02 *(0.02)* | 0.14 *(0.12)* | 0.56 *(0.03)* | 56.25 *(0.51)* | 0.26 *(0.04)* | 0.02 *(0.00)* | 99.27 |
| #16 | Glass (21%) | 53.91 *(0.25)* | 0.67 *(0.04)* | 15.22 *(0.65)* | 1.44 *(0.03)* | 15.08 *(1.08)* | 12.29 *(0.57)* | 0.21 *(0.02)* | 98.82 |
| | Olivine (32%) | 41.97 *0.17)* | 0.02 *(0.01)* | 0.07 *(0.01)* | 1.69 *(0.01)* | 55.45 *(0.22)* | 0.21 *(0.01)* | 0.18 *(0.02)* | 99.59 |
| | Orthopyroxene (45%) | 57.62 *(0.33)* | 0.08 *(0.01)* | 1.21 *(0.11)* | 1.01 *(0.04)* | 38.28 *(0.30)* | 0.49 *(0.01)* | 0.64 *(0.07)* | 99.32 |
| #17a (using 4) | Glass (93%) | 18.67 *(0.20)* | 2.36 *(0.16)* | 45.04 *(0.54)* | 0.01 *(0.01)* | 1.34 *(0.11)* | 31.77 *(0.74)* | 0.00 *(0.00)* | 99.19 |
| | Gehlenite (7%) | 21.36 *(0.08)* | 0.15 *(0.02)* | 36.67 *(0.18)* | 0.01 *(0.01)* | 0.09 *(0.01)* | 41.27 *(0.23)* | 0.00 *(0.00)* | 99.55 |
| #17b (using 6) | Glass (48%) | 19.46 *(0.21)* | 2.97 *(0.20)* | 49.27 *(0.73)* | 0.01 *(0.01)* | 2.01 *(0.13)* | 25.35 *(0.79)* | 0.00 *(0.00)* | 99.07 |
| | Gehlenite (52%) | 20.97 *(0.08)* | 0.13 *(0.02)* | 36.42 *(0.09)* | 0.02 *(0.01)* | 0.09 *(0.01)* | 41.16 *(0.31)* | 0.00 *(0.00)* | 98.79 |
| #18 | Glass (84%) | 5.94 *(0.06)* | 2.03 *(0.02)* | 55.33 *(0.17)* | 0.01 *(0.01)* | 0.20 *(0.01)* | 35.75 *(0.16)* | 0.00 *(0.00)* | 99.26 |
| | Grossite (16%) | 0.34 *(0.03)* | 0.08 *(0.01)* | 76.67 *(0.18)* | 0.00 *(0.00)* | 0.07 *(0.02)* | 22.43 *(0.05)* | 0.00 *(0.00)* | 99.59 |
| #20 (using 15) | Glass (26%) | 53.74 *(0.12)* | 0.98 *(0.02)* | 16.03 *(0.04)* | 0.62 *(0.00)* | 13.24 *(0.03)* | 14.13 *(0.01)* | 0.02 *(0.00)* | 98.76 |
| | Olivine (74%) | 41.25 *(0.20)* | 0.01 *(0.00)* | 0.07 *(0.01)* | 0.73 *(0.03)* | 56.55 *(0.25)* | 0.24 *(0.02)* | 0.01 *(0.00)* | 98.85 |
| #21 (using 15) | Glass (69%) | 47.02 *(0.16)* | 0.89 *(0.02)* | 18.67 *(0.11)* | 1.08 *(0.02)* | 19.75 *(0.01)* | 11.49 *(0.01)* | 0.02 *(0.01)* | 98.92 |
| | Olivine (31%) | 42.44 *(0.35)* | 0.02 *(0.01)* | 0.12 *(0.03)* | 0.80 *(0.01)* | 56.56 *(0.03)* | 0.29 *(0.00)* | 0.02 *(0.00)* | 100.25 |
| #22 (using 16) | Glass (26%) | 41.20 *(0.29)* | 0.19 *(0.04)* | 22.93 *(0.44)* | 0.67 *(0.04)* | 18.80 *(0.34)* | 13.51 *(0.16)* | 0.35 *(0.12)* | 97.66 |
| | Olivine (50%) | 41.07 *(0.26)* | 0.09 *(0.04)* | 0.75 *(0.43)* | 1.67 *(0.01)* | 54.91 *(0.29)* | 0.45 *(0.11)* | 0.24 *(0.08)* | 99.19 |
| | Orthopyroxene (23%) | 59.31 *(0.48)* | 0.05 *(0.02)* | 0.73 *(0.14)* | 0.57 *(0.02)* | 39.74 *(0.16)* | 0.31 *(0.05)* | 0.02 *(0.01)* | 100.73 |
| | Al-Spinel (1%) | 0.34 *(0.02)* | 0.08 *(0.01)* | 70.24 *(0.29)* | 0.02 *(0.00)* | 28.10 *(0.18)* | 0.62 *(0.12)* | 0.90 *(0.02)* | 100.24 |

**Table 5.** Average residual glass and phase compositions (wt%) for reduced runs from Pt-loop equilibrium crystallization experiments.

| Experiment | Phase/Abundance | SiO$_2$ | TiO$_2$ | Al$_2$O$_3$ | FeO | MgO | CaO | Cr$_2$O$_3$ | Total |
|---|---|---|---|---|---|---|---|---|---|
| #6-1r (reduced fO$_2$) | Glass (42 %) | 22.52 *(0.12)* | 0.10 *(0.02)* | 37.00 *(0.08)* | 0.02 *(0.01)* | 0.13 *(0.01)* | 41.77 *(0.08)* | 0.01 *(0.00)* | 101.53 |
| | Grossite (8 %) | 0.83 *(0.13)* | 0.07 *(0.01)* | 76.77 *(0.52)* | 0.01 *(0.00)* | 0.36 *(0.01)* | 22.09 *(0.43)* | 0.00 *(0.00)* | 100.12 |
| | Gehlenite (50 %) | 22.31 *(0.02)* | 1.73 *(0.06)* | 40.49 *(0.10)* | 0.02 *(0.00)* | 0.18 *(0.01)* | 36.46 *(0.12)* | 0.00 *(0.00)* | 101.19 |
| #1r (reduced fO$_2$) | Glass (92%) | 28.29 *(0.21)* | 1.13 *(0.04)* | 33.56 *(0.44)* | 0.01 *(0.01)* | 5.14 *(0.03)* | 31.08 *(0.03)* | 0.01 *(0.01)* | 99.22 |
| | Spinel (8%) | 0.75 *(0.16)* | 0.07 *(0.02)* | 71.66 *(0.16)* | 0.00 *(0.00)* | 27.35 *(0.31)* | 1.10 *(0.12)* | 0.19 *(0.53)* | 101.12 |

Orthopyroxene stability was confirmed by experiments in #14, #16, and #22. Experiments at 1510K (#22) and 1540K (#14) at 50x dust enrichment produced similar abundances and chemistry of forsteritic olivine (48-50%), orthopyroxene (25-23%), and Al-spinel (1%). As the system shifts towards higher dust enrichment at constant temperature (from #14 at 50x to #16 at 100x), the modal abundance of olivine decreases (48% to 32%) while orthopyroxene (25% to 45%) increases. Also, the system stabilizes Cr-spinel rather than Al-spinel as noted in Table 3 and the Mg-Ca-Al map (Fig. 4g, h). Cr-spinel analyses of #16 are not included in Table 4 because the fine sizes of crystals made them impossible to analyze by EPMA without overlapping surrounding glass. Experimental and predicted abundances of both olivine and orthopyroxene are found to be in very good agreement (Table 3).

The effect of $fO_2$ on experimental results was assessed by running compositions #1 and #6-1 at IW-3.4, approximately the $fO_2$ of a solar composition vapor enriched 50x in CI-like dust (Fig. 5). Results (#1r and #6-1r of Table 5) are nearly identical to those for the same compositions at FMQ (#1 and #6.1 of Table 4). Neither Ca-rich pyroxene nor perovskite was observed in the products of either of these experiments.

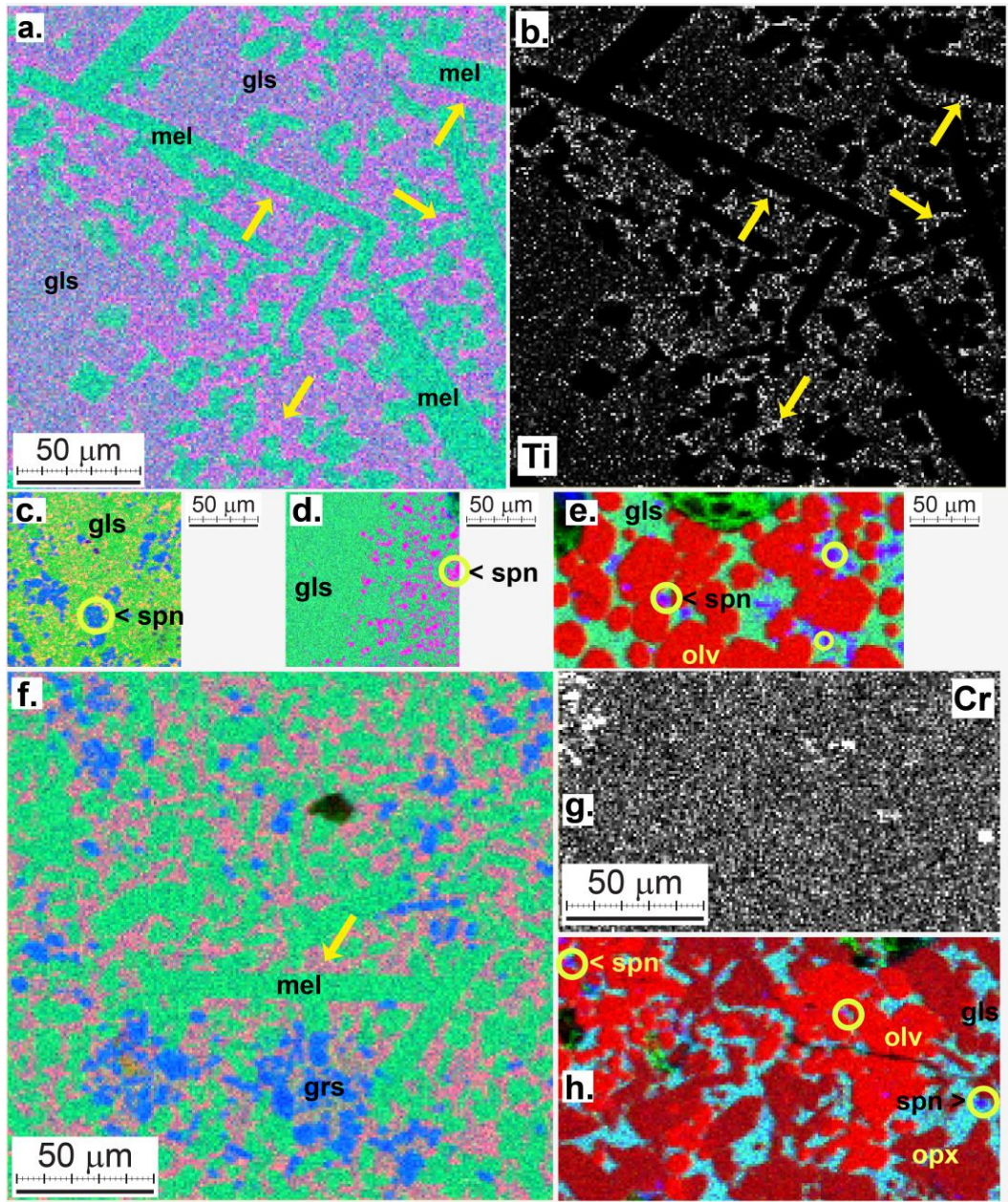

**Fig. 4.** X-ray intensity maps (all 1 μm/pixel) combined into composite red-green-blue (RGB) three element images: **a.** #17a (TiCaAl), **b.** #17a, Ti only, contrast enhanced, **c.** #7 (TiCaAl), d. #5b (MgCaAl), **e.** #11-1 (MgCaAl), and **f.** #6-1 (TiCaAl), **g.** #16 (Cr only), **h.** #16 (MgCaAl). Yellow arrows indicate Ti buildup in mesostasis (gls) at the edges of growing gehlenitic melilite (mel) crystals. Yellow circles identify Al-spinel (in **c-e**) and Cr-spinel rich (in **h**; **g** is the same area in Cr intensity). Labels indicate olv (Mg-olivine), grs (grossite), and opx (Mg-pyroxene).

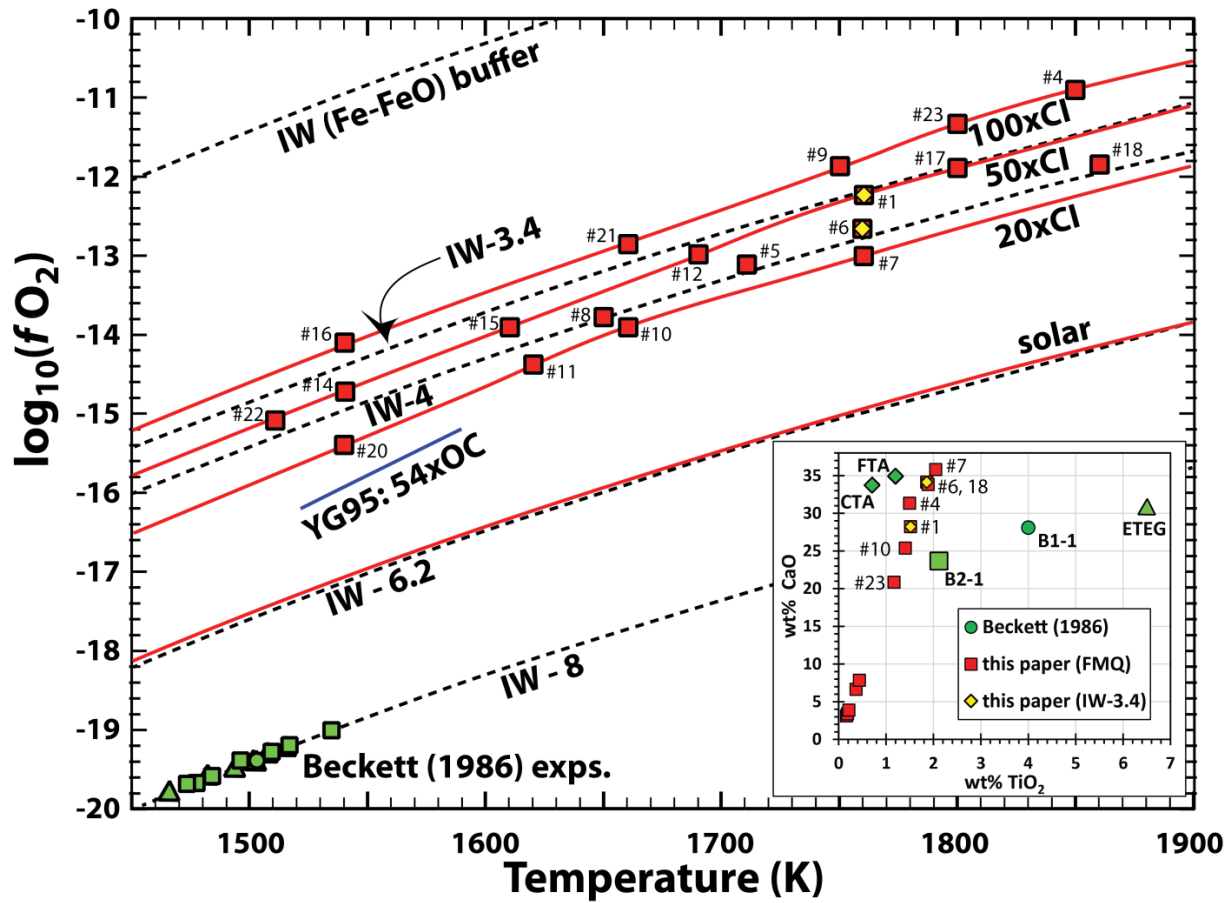

**Fig. 5.** Oxygen fugacity and temperatures predicted for solar and CI-dust enriched systems (Ebel and Grossman, 2000; solid red curves) and IW-$n$ buffer curves (dashed). Low $f(O_2)$ experiments of Beckett (1986) are plotted (green symbols: ETEG triangles, B1-1 circle, B2-1 squares) at reported $f(O_2)$ and T. Squares indicate T and dust enrichment of the present experiments (numbers as in Fig 3a, Tables 1, 2), at their predicted $f(O_2)$ (Ebel and Grossman, 2000). A trajectory for vapor enriched 54x in ordinary chondrite dust (Yoneda and Grossman, 1995) is provided for comparison. Inset compares present (Table 1 only) CaO - $TiO_2$ compositions for the experiments Beckett (1986).

## 4. DISCUSSION

Perovskite did not grow in any high-T silicate liquid, even in runs where Ti built up, but Ca was depleted, outside the rims of growing gehlenite crystals (Fig. 4a,b; 4f). In the $TiO_2$-bearing experiments, Ti remains in the liquid. This result is consistent with the absence of perovskite in most igneous (melted; Type B) CAIs (Grossman 1980). Perovskite is observed to be common in non-igneous "fluffy" CAIs, and is, indeed, predicted to form from vapor at high T and $d < \sim15x$ (Fig. 1). Titanium is as refractory as Al and Ca, so predictions that nearly all Ti condenses above the stability field of olivine are likely correct at all $d$. Therefore it is not an absence of $TiO_2$ in the condensed assemblage that prevents growth of perovskite in the experiments. Our experiments allow us to conclude that perovskite is not a stable phase in dust-enriched systems in which liquid is present at high T, and where the ratios of major refractory elements Al, Ca, and Ti are solar.

In contrast, Beckett (1986) performed experiments that constrained the stability conditions of perovskite and Al-, Ti-rich calcic pyroxene (a.k.a. fassaite) in bulk compositions analogous to those he measured for particular CAIs. In Fig. 5 we plot for comparison the wt% CaO and $TiO_2$ of his experiments relative to ours which follow the solar Ca/Ti trend. He ran compositions CTA and FTA only in air, and B2-1, B1-1, and ETEG in air and at IW-8. His FTA and CTA crystallized Prv after Mel and Spn, at 1725 and 1637 K, respectively, well below our #4 (1860 K) or #6 (1760 K). His ETEG, in air, crystallized Prv at 1508 K, after An. He found that $Ti^{3+}$ stabilized the pyroxene structure at IW-8, with liquid lines of descent B2-1 (2.1 wt% $TiO_2$) An > Spn > Cpx > Mel; B1-1 (4 wt% $TiO_2$) Cpx > Mel > Spn > An; and ETEG (6.5 wt% $TiO_2$) Cpx > Mel > An > Prv > Wo. Abbreviations are: An=anorthite, Spn=Al-spinel, Cpx=Al-, Ti-, Ca-rich pyroxene, Mel=melilite, Prv=perovskite, and Wo=pseudowollastonite.

The purpose and outcomes of the present experiments differ from those of Beckett (1986) in bulk compositions, temperatures, and $fO_2$. We chose compositions #1 and #6 for low $fO_2$ runs because their predicted condensate bulk compositions, just above the stability field of olivine (Fig. 3), should be closest to a pyroxene stability field, since their MgO and $SiO_2$ contents are higher than other points above the olivine stability field and they have high $TiO_2$ contents (Fig. 5, inset). Olivine is rarely observed with Al-, Ti-rich Ca-pyroxene in meteoritic CAIs. Our confirmation of the lack of either perovskite or pyroxene in these experiments (Fig. 3b) confirms only that neither phase is stable in CI-dust enriched vapor at the temperatures we investigated. We do not contradict the conclusion of Beckett (1986) that Ti-rich pyroxene (and not perovskite) is stabilized in $TiO_2$-rich melts by oxygen fugacities much lower than that calculated for a gas of solar composition.

Spinel stability is more puzzling. Spinel is seen in experimental products where it is predicted above the temperature of olivine stability, but only experiment #11 has spinel crystallizing with olivine. Experiment #20 is ambiguous for reasons detailed above, so the spinel + olivine + liquid field may extend to higher dust enrichments, perhaps even including point #8 (Fig. 3b). Appearance of Al-spinel with the growth of orthopyroxene (enstatite) from olivine (cf., Ebel and Grossman, 2000) at significantly lower temperatures with increased condensation of $SiO_{(g)}$ into liquid can be understood as due to the availability of MgO in the liquid. However, Al-rich spinel is *not* predicted to occur with orthopyroxene (Fig. 3a). The absence of Al-spinel in much of the liquid + olivine field is consistent with the observed rarity of spinel in ferromagnesian chondrules. Consistent with predictions, Cr-rich spinel appears only in experiment #16.

The finding of grossite ($CaAl_4O_7$, #7, #18) as predicted, is gratifying. The stability of melilite (near-pure gehlenite, $Ca_2Al_2SiO_7$) with grossite in #6 was verified directly and by proxy using composition #4. The melilite field extends to #17, but not to #4. This difference between experiment and prediction is likely due to the inadequacy of the melilite solid solution model used in the prediction, which is demonstrably inadequate (Grossman et al., 2002, their Fig. 1). These experiments confirm that CI dust enrichments up to nearly 100x are, contrary to predictions, entirely consistent with the ubiquity of melilite in igneous Type B CAIs (Grossman, 1980).

These results demonstrate the importance of experimental testing of predictions that are based on thermodynamic models, even when the models are based on sound, internally consistent equations of state. Generally, the predictions using the MELTS model were found to match experimental phase relations and abundances better than the predictions using the CMAS model. The experiments serve to expand the scope of tested crystal-liquid equilibrium pairs (e.g., LEPR, 2013) that may be used to calibrate and/or test future models for liquid-crystal equilibration, particularly in high-T silica-poor compositions.

## ACKNOWLEDGEMENTS

The authors are grateful for constructive reviews by L. Grossman, A. Davis and E. Bullock, that significantly strengthened and improved the manuscript by suggesting experiments at low $fO_2$. Associate editor Chris Herd handled the manuscript. The authors thank Shawn Wallace (AMNH) and Hanna Nekvasil (Stony Brook) for analytical and laboratory assistance. This work was supported by U.S. N.A.S.A. grant NNX10AI42G (DSE) and an AMNH Katherine Davis post-doctoral fellowship (GU). This research has made use of NASA's Astrophysics Data System Bibliographic Services.